# EPR and Non locality

## Understood in terms of Kuramoto Synchronisation

A. TAMBYRAJAH

Principal Lecturer*, School of Computing and Mathematical Sciences,
University of Greenwich, London SE10 9LS, UK,
email:atambyrajah@yahoo.co.uk

ABSTRACT. *A new way of understanding the non local correlation effects observed in the "Twin photon" EPR experiment is presented. The theory is based on a vector version of the Kuramoto synchronisation model for coupled nonlinear oscillators. EPR correlation is obtained without the use of superluminal signals. Bell's inequality is used to confirm that the local theory developed can still violate the inequality resolving the non local nature of the correlation.*

## 1 Introduction

Bell's paper [1] and the related Bell's inequality represents a seminal work on the nature of Quantum Mechanics. It provides a mathematical frame work for considering local hidden variable theories particularly in the explanation of the EPR experiment. The conclusion of the paper that all local hidden variable theories produce correlation results not compatible with quantum theory still remains valid. The leading conclusion that the correlation in EPR is a non-local effect has raised many questions on the full implication of the assumptions that underlie Bell's theorem. In this work, in contrast to Bell, we shall follow arguments proposed by Bohm [2] in developing a local theory that can still explain the non local effect yet maintaining the causal restrictions of relativity.

In Bohm [2], a possibility of explaining non locality using nonlinear coupling within a local field is outlined. Bohm argues that Einstein was not too concerned by the non-locality predicted by Quantum mechanics but rather the expectation was that a new broader theoretical framework could provide a basis of more fundamental notion of reality that is local and relevant. Bohm draws attention to the fact that nonlinear equations have the property that small deviations can lead to rapid oscillatory motion around limit cycles. This leads for example to coupled-nonlinear oscillators producing synchronisation or anti synchronisation even when separated in space and time.

Synchronisation of nonlinear oscillators is a well known phenomenon in physics. As early as 1665, Huygens observed that his pendulums that were separated by some distance performed sympathetic oscillations even though there were no apparent connections between them. More recently, in the quantum mechanical context, the Josephson Effect is attributed to synchronisation of quantum waves that weakly overlap with each producing correlated pairs of electrons.

The Josephson junction effect is explained by Feynman [3] in terms of weakly coupled phase coherent systems. Each system is characterised by a single quantum wave. The waves overlap each other slightly without disturbing each other to produce a weak coupling. The strength of the supercurrent produced depended on the relative phases of the waves on either side of the barrier. If the phases could somehow be driven slightly out of step, the supercurrent would turn on within the two conductors. The supercurrent was predicted to be proportional to the **_Sine_** of the phase difference. The phases are driven out of step using an external source of current fed to the barrier system. The maximum current is obtained when the phase difference is π/2. Feynman further suggested the Josephson type effect can occur for any pair of phase coherent systems coupled by any sort of weak link. However in the case of EPR experiment a question remains as to whether a local theory can form a basis for a non local effect. Bohm's argument for an extension of the Bell's correlation function provides an explanation for this possibility.

## 2  Bohm's extension of Bell's theorem

In the Bell EPR model, each experimental result is considered to be completely determined by a set of hidden variables λ. Bohm as with Bell assumes that the result of measurement A of the spin direction â is determined only by λ and â, while the result of measurement B of the spin direction b̂ is determined only on b̂ and λ. Importantly, Bohm brings to attention that in Bell's theorem the hidden variables do not have any imposed restrictions on the nature of their dynamics. This means that they can be non local if required. The only restriction is that the response of the observing instrument to the set λ depends only on its own state and **not** on the state of any other piece of apparatus far away. Thus, A= A (â, λ) and B = B (b̂, λ), with the possibilities A= A (â, b̂, λ) and B= B (â, b̂, λ) being excluded.

Bell's final assumption is that the statistical distribution of hidden variables can be given by a density function ρ (λ) which is independent of â and b̂. The correlation function P (**â**, b̂) is then defined in terms the density function as

$$P(\hat{a}, \hat{b}) = \int \rho(\lambda) A(\hat{a}, \lambda) B(\hat{b}, \lambda) d\lambda$$

From this definition, the CHSH [4] form of Bell's inequality can be obtained to give
$$|P(\hat{a},\hat{b}) - P(\hat{a},\hat{d})| + |P(\hat{c},\hat{d}) + P(\hat{c},\hat{b})| \leq 2$$

In Bohm's extension of Bell's inequality, the density function dependency includes the measurements **â** and b̂ as well as the hidden variable λ giving the functional relationship ρ=ρ (**â**, b̂, λ). This does not imply that **â *and*** b̂ interact with each other or with λ. He illustrates this by making comparison with the thermodynamic properties of Pressure (P) and Temperature (T) of an aggregate collection of atoms being not a separate property but rather an abstraction from the underlying atomic variables α, the probability distribution over α is argued to take the form ρ=ρ (α, T, P ...). He comments that this does not mean T and P "interact" with each other or with α

Bohm [2] summarises role of hidden variables in EPR as follows,

> *"The distribution of the total set of hidden variables is restricted in such a way that whenever it gives rise to the emergence of given orientations of two pieces of apparatus, it gives rise also to the emergence of properly coordinated pairs of particle spins"*

With density function set to be ρ=ρ (â, b̂, λ), correlation function now becomes

$$P(\hat{a}, \hat{b}) = \int \rho(\lambda) A(\hat{a}, \lambda) B(\hat{b}, \lambda) d\lambda$$

The Bell-CHSH [4] derivation of the inequality also follows from above correlation function.

Using the hidden variables of the actual positions of all particles consisting of the observed object plus the observing apparatus, Bohm shows that Bell's inequality is violated demonstrating that his non relativistic localised quantum potential, based on single wave function for the entire two particle system, can provide an explanation of the observed non local effect to be consistent with quantum mechanics. Given this, we now consider a new local model to explain the non local effect present in the twin photon EPR experiment.

### 3   The Nonlinear synchronisation "twin photon" EPR Model

A recent experiment for the Twin Photon EPR is described in G. Weihs [5]. The single source for the twin photons in this experiment is provided by a Parametric Down-Conversion process. The correlation of the plane polarisation between the two 700 nm photon beams is measured by a complex electronic coincidence process for analysers at various angles.

A limited schematic diagram for the experiment relevant for our model is shown below.

Twin linearly polarised photon pairs

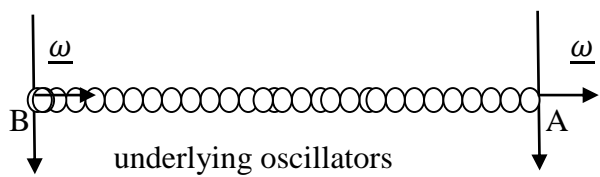

**Fig 1 Schematic diagram for the twin pair photon EPR experiment**

The correlation in our model is postulated occur through vector synchronisation of the underlying vacuum oscillators generated by photons at A and B.

Further, there are other oscillations, referred to as 'noise', produced by background particles and photons. The actual cause and process for the presence of these oscillations or even the actual nature of the oscillations need not be considered at this stage except to postulate that there are oscillations generated with angular frequency vectors dependent on the angular frequency vectors of the photons present.

The correlation expected in the model relies on the process of synchronisation of coupled nonlinear oscillators. As already described, there are many examples of this type of synchronisation in physics and also other areas, Strogatz [6]. The synchronisation process in the alignment of long range magnetisation of superfluids [3] is a relevant example for our proposed model. In this case there is not only a synchronisation of spin (frequency in our case)) values but also the alignment of spin directions. The synchronisation process is also found to be fast. It is this type of vector alignment that we seek to use in the consideration of the correlation of polarisation in the twin pair EPR effect

The description presented below relies *on a phasor diagram view* of the oscillation as a circular motion with a given angular frequency $\boldsymbol{\omega}$. The angular frequency is represented as a vector, $\underline{\omega}$ perpendicular to the plane of the circular motion in the phasor representation. The plane of the circular representation is the plane of polarisation of the photons.

In fig (2), the twin pair photons at A and B are shown to have equal frequencies $\omega$. The index label $\boldsymbol{i}$ are for oscillators generated at position B. The index label $j$ are for oscillators generated from position A. In this model, the synchronisation of the coupled oscillators in the line from A to B are considered.

The angular frequency vector $\underline{\omega}_j$, is shown perpendicular to the plane of the polarisation. The effect of generating a new polarisation direction by the analyser at A can be represented by a rotation of the vector $\underline{\omega}$ about the axis of travel by the use of a rotation operator $L_j(\theta')$. The effect of rotation $L_j(\theta')$ represents change of polarisation direction by an angle, $\theta'$. The new polarisation direction after the rotation is shown by the vector $\underline{\Omega}_j$. The synchronisation process results in the oscillators generated from A being rotated with their angular velocity vector to $\underline{\Omega}_i$. This in turn is postulated to change photon polarisation at A to perpendicular to the vector $\underline{\Omega}_i$.

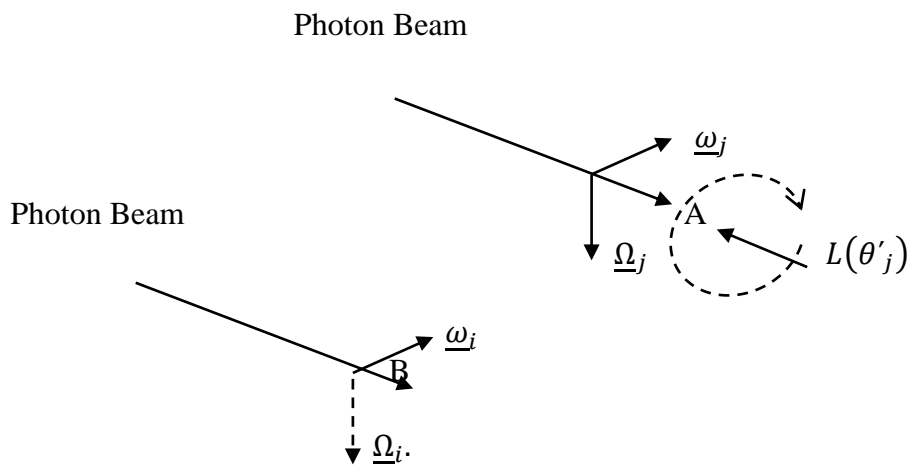

Fig 2 Twin pair photons with initial vertical polarisation

The proposed model for the synchronisation is based on a vector version of the Kuramoto [7] synchronisation model. It is described by equation (1) where $\hat{\omega}$ is the unit vector in the direction of angular frequency vector $\underline{\omega}$

$$\underline{\Omega}_i = L(\theta'_N)\,\underline{\omega}_i + \frac{K}{N}\sum_{j=1}^{N} Sin(\theta_j - \theta_i) L(\theta'_j)\hat{\omega}_j \qquad (1)$$

With $L(\theta'_1) = L(\theta')$, $L(\theta'_2) = L(\theta' + \epsilon\theta')$, ...... $L(\theta'_3) = L(\theta' + \epsilon\theta' + \epsilon^2\theta')$, $L(\theta'_N) = L(\theta' + \epsilon\theta' + \epsilon^2\theta' + \cdots \ldots \ldots \epsilon^{N-1}\theta')$

Using established procedures described in Strogatz [8], we can approach a solution for the above vector version of the Kuramoto [7] model.

From the definitions given below, the summation in (1) can be simplified to include the order parameter term r, through

$$re^{i\phi}\overline{L}(\theta')\overline{\hat{\omega}} = \frac{1}{N}\sum_{j=1}^{N} e^{i\theta_j} L(\theta'_j)\hat{\omega}_j \qquad (2)$$

The parameter r reflects the degree of coherence in the system and $\overline{\hat{\omega}}$ is the unit vector for the mean of the unit angular frequency vectors $\hat{\omega}_j$.

As the numbers oscillators N tends to infinity, the summation in (2) can be replaced to an integral using the following definitions.

$$re^{i\phi} = \int_0^{2\pi} e^{i\theta} \left(\frac{1}{N}\sum_{j=1}^{N} \delta(\theta - \theta_j)\right) d\theta \qquad (3)$$

$$re^{i\phi} = \int_0^{2\pi} \int_{-\infty}^{+\infty} e^{i\theta} \rho(\theta, \Omega, \omega, t) g(\omega) d\omega d\theta \qquad (4)$$

Where, $\phi$ is the mean phase of all oscillators involved in the synchronisation process, $g(\omega)$, is the distribution function for oscillator frequencies. In our case a $\delta(\omega)$ distribution will be used. ,$\Omega_i$., represents the density for each ω at a time t and phase angle, $\Omega_i$. that is normalised to 1 over a phase of $2\pi$ for the continuum of oscillators.

$$\int_0^{2\pi} \rho(\theta, \Omega, \omega, t) d\theta = 1 \qquad (4b)$$

Incorporating these definitions, equation (1) can be written as,

$$\underline{\Omega}_i = L(\theta'_N)\underline{\omega}_i + Kr\, Sin(\phi - \theta_i)\overline{L}(\theta')\overline{\hat{\omega}} \qquad (5)$$

Where $\overline{L}(\theta') = L(\theta' + \epsilon\theta')$ and $L(\theta'_N) = L(\theta' + \epsilon\theta')$, to the first order in $\epsilon$.
In seeking a first look at a solution to our synchronisation equation (5), we use the following simplifying assumptions,

$$\overline{L}(\theta') \approx L(\theta'), \; L(\theta'_N) \approx L(\theta')$$

Using $\underline{\omega}_i = \omega\hat{\overline{\omega}}$ and $L(\theta')\hat{\overline{\omega}} = \underline{\Omega}$, equation (5) can be reduced to

$$\underline{\Omega}_i = \omega_i\hat{\underline{\Omega}} + Kr\, Sin(\varphi - \theta_i)\hat{\underline{\Omega}} \qquad (6)$$

Taking account of the vector directions, the following scalar equation is valid.

$$\Omega_i = \omega_i + Kr\, Sin(\varphi - \theta_i) \qquad (7)$$

This is a form of the Kuramoto equation [7] modified for infinite number of oscillators with also the inclusion of the order parameter r. We use established procedures to solve this equation. The approach of Strogatz and Mirollo [8] will be presented within the context of the twin photon correlation problem.

## 4  Incoherent Solution

The incoherent solution corresponds to a state in which for each $\omega$, all the photon generated oscillators are uniformly distributed around the phase circle. The incoherent solution $\rho_\circ(\theta, t, \omega)$ = $1/2\pi$ was obtained by Kuramoto [7] and further confirmed by Strogatz [8] for all, $\theta, t$ and $\omega$. The density function $\rho(\theta, t, \omega)$ was defined to be a normalised quantity. It is also required to be a conserved quantity and therefore satisfying the continuity equation,

$$\frac{\partial \rho}{\partial t} = -\frac{\partial(\rho v)}{\partial \theta} \qquad (8)$$

Where 
$$v = \omega + Kr Sin(\emptyset - \theta) \qquad (8b)$$

is considered the drift angular frequency.

Substitution of $\rho_\circ(\theta, t, \omega) = 1/2\pi$ into equation (4) shows

$$re^{i\emptyset} = \int_0^{2\pi} \int e^{i\theta} \left(\frac{1}{2\pi}\right) g(\omega) d\omega d\theta = 0 \qquad (9)$$

This gives from 8(b), $v(\theta, t, \omega) = \omega$ and hence independent of $\theta$. The continuity equation is therefore satisfied with,

$$\frac{\partial \rho}{\partial t} = 0 \text{ and } \frac{\partial(\rho v)}{\partial \theta} = 0$$

Thus the density of $1/2\pi$ is an incoherent solution satisfying the Kuramoto equation. The incoherent solution of $1/2\pi$ also forms a solution to the proposed photon correlation equation (1). It is further found that small fluctuations in the mean density around $1/2\pi$ does lead to first order and higher order solutions, providing the possibility of synchronised or anti-synchronised solutions in the correlation of photon polarisation.

## 5  First Order Solutions

In searching for the first order solution for the coherence r, we assume a fluctuation of the mean density of the following form

$$\rho(\theta, \Omega, \omega, t) = 1/2\pi + \varepsilon \eta(\theta, \Omega, \omega, t) \qquad \varepsilon \ll 1 \quad (10)$$

From the normalisation condition (4b),

$$\int_0^{2\pi} \eta(\theta, \Omega, \omega)\, d\theta = 0 \text{ for all } \theta, t \text{ and } \omega \quad (10b)$$

The continuity equation (8) is now,

$$\varepsilon \frac{\partial \eta}{\partial t} = -\frac{\partial \left[\left(\frac{1}{2\pi} + \varepsilon \eta\right) v\right]}{\partial \theta} \quad (11)$$

In order to isolate the first order solution r (t) is separated into two parts,
r (t) = $\varepsilon r_1$ (t) + O ($\varepsilon^2$), $r_1$ of order $\varepsilon$ and the remaining higher orders as O ($\varepsilon^2$). Substitution in (9) defines $r_1$ (t) as

$$r_1(t)\, e^{i\phi} = \int_0^{2\pi} \int e^{i\theta}\, \eta(\theta, \Omega, \omega, t) g(\omega)\, d\omega\, d\theta \quad (9b)$$

The continuity equation (8), to order $\varepsilon$, is now,

$$\frac{\partial v}{\partial \theta} = -\varepsilon r_1 \cos(\phi - \theta)$$

The density evolution equation (11) therefore becomes

$$\varepsilon \frac{\partial \eta}{\partial t} = -\varepsilon \omega \frac{\partial \eta}{\partial \theta} + \varepsilon r_1 \cos(\phi - \theta) \quad (12)$$

As $\eta(\theta, t, \omega)$ is real and $2\pi$ periodic in $\theta$, Fourier methods are used to analyse the solutions to (12),

$$\eta(\theta, \Omega, \omega, t) = c(\Omega, \omega, t) e^{+i\theta} + c^*(\Omega, \omega, t) e^{-i\theta} + \eta^{\perp}(\theta, \Omega, \omega, t) \quad (13)$$

The first two terms on RHS represents the first harmonic in the solution and the third term $\eta^{\perp}(\theta, \Omega, \omega)$ denotes all other higher order terms. Equation (13) has zero mean. First harmonic is the only term that contributes to the coherence r (t). The higher harmonic term $\eta^{\perp}$ makes no contribution to r (t). This can be seen as follows
Writing $r_1 \cos(\phi - \theta)$ as $Re\, [r_1 e^{i\phi} e^{-i\theta}]$, and also substituting (13) into 9(b) we have

$$r_1(t)\, e^{i\phi} = 2\pi \int_{-\infty}^{+\infty} c^*(\Omega, \omega, t)\, g(\omega)\, d\omega$$

It follows, 
$$r_1 \cos(\phi - \theta) = Re\left[2\pi \int_{-\infty}^{+\infty} c^*(\Omega, \omega, t)\, g(\omega)\, d\omega\, e^{-i\theta}\right]$$

$$r_1 \cos(\phi - \theta) = \pi \int_{-\infty}^{+\infty} c(\Omega, \omega, t)\, g(\omega)\, d\omega\, e^{+i\theta} + \text{c.c. of first term}$$

(14)

The result shows that the coherence term $r_1$ depends only on $c(\Omega,\omega,t)$ and it's complex conjugate.

## 6 First Harmonic

Using the Fourier analytic form (13) for the solution, we look at the fundamental coherence mode

The solution (13)
$$\eta(\theta,\Omega,\omega,t) = c(\Omega,\omega,t)e^{+i\theta} + c^*(\Omega,\omega,t)e^{-i\theta} + \eta^\perp(\theta,\Omega,t,\omega)$$

And the result, $r_1 \cos(\phi - \theta) = \pi \int_{-\infty}^{+\infty} c(\Omega,\omega,t)\, g(\omega)d\omega\, e^{+i\theta}$ + c.c., is substituted into the equation

$$\frac{\partial \eta}{\partial t} = -\varepsilon\omega\frac{\partial \eta}{\partial \theta} + \varepsilon K r_1 \cos(\phi - \theta).$$

Equating coefficients of exponential terms on both sides, we have
$$\frac{\partial c}{\partial t} = -i\omega c + \frac{K}{2}\int_{-\infty}^{+\infty} c(\Omega,\nu,t)g(\nu)d\nu \qquad (15)$$

The time dependency of c is obtained from solving the above equation. The coherence $r_1$ and its time dependence can then be determined.

## 7 Coherence Spectrum

A solution of the form, $c(\Omega,\omega,t) = b(\omega)e^{\Omega t}$ is first considered. Substituting into (15) we have

$$\Omega b = -i\omega b + \frac{K}{2}\int_{-\infty}^{+\infty} b(\nu)g(\nu)d\nu \qquad (16)$$

The integral in (15) from the definition of $r_1 e^{i\phi}$ is some constant. Writing

$$\frac{K}{2}\int_{-\infty}^{+\infty} b(\nu)g(\nu)d\nu = A \qquad (17)$$

Therefore from (16) it follows

$$b(\omega) = \frac{A}{\Omega + i\omega} \qquad (18)$$

This implies that $b(\nu)$ must also be of the same form. Substituting (18) into (16) we have that either A=0, or

$$\frac{K}{2}\int_{-\infty}^{+\infty}\frac{A}{\Omega + i\nu}g(\nu)d\nu = A$$

Giving

$$\frac{K}{2}\int_{-\infty}^{+\infty} \frac{g(v)}{\Omega+iv} dv = 1 \qquad (19)$$

Equation (19) is the equation for the discrete spectrum. Case A=0 leads to c ($\omega$)=0 for all $\omega$.

We now seek a solution for our twin photon synchronisation problem under the conditions of (1) a noiseless background and (2) a delta function distribution for the distribution, $g(\omega)$.
If the frequency distribution is an even function
  i.e. $g(\omega) = g(-\omega)$,
and it is non increasing on [0,∞],
  i.e. $g(\omega) \leq g(v)$ for all $\omega \geq v$,
then at most one solution exists for equation (19) and it is a real solution, Mirollo [9].

Separating the real and imaginary parts in (19) we have

$$1 = \frac{K}{2}\int_{-\infty}^{+\infty} \frac{\Omega}{\Omega^2+v^2} g(v)dv - \int_{-\infty}^{+\infty} \frac{iv}{\Omega^2+v^2} g(v)dv \qquad (20)$$

In Theorem 2, Mirollo [9], it is shown that for any non-increasing even function such as, $g(\omega') = \delta(\omega')$, the imaginary part of integral in (20) will tend to zero.
With $\omega' = v - \omega_1$, we have the real part of the integral in (20) as

$$1 = \frac{K}{2}\frac{\Omega}{\Omega^2+\omega_1^2} \qquad (21)$$

To determine the resultant correlation of the polarisation of the twin photons, we need first look at the coherence term $r_1$ determined by the function $c(\Omega, \omega, t) = b(\omega)e^{\Omega t}$

Solving the quadratic equation and setting the quadrant, $K^2/4 - 4\omega_1^2$ equal to zero to give the one allowed real solution.
We have the resultant value of the frequency $\Omega$ at B, $\omega_i = \omega_1$.

From (18)

$$b(\omega_1) = \frac{1}{\omega_1+i\omega_1} = \frac{1}{2\omega_1} - \frac{i}{2\omega_1} \qquad (22)$$

Now from (14)
$r_1 \cos(\phi - \theta) = \pi \int_{-\infty}^{+\infty} c(\Omega, \omega) g(\omega)d\omega \, e^{+i\theta}$ + the complex conjugate(c.c.) of the first term.
Including the frequency distribution $g(\omega)$ function to be $\delta(\omega_1)$ and incorporating the result (22) for $b(\omega_1)$ we have

$$r_1 \cos(\phi - \theta) = \pi e^{\Omega t} \frac{1}{2\omega_1} \sqrt{5} \, Sin(\theta + \alpha) \qquad (23)$$

The mean phase $\phi$ for the oscillators can be reasonably taken to be closely following the originating phase $\theta$. If we consider case $\phi = \theta$ together with the result $\Omega = \omega_1$, we can obtain a picture of the coherence varying with time, as

$$r_1 = \pi e^{\omega_1 t} \frac{1}{2\omega_1}\sqrt{5} Sin(\omega_1 t + \alpha)$$

The positive exponential term in the solution for $r_1$ means that for laser frequency values of the order for example of $10^{15}$ Hz, $r_1$ does oscillate its way to infinity in a very short time. This is described as unstable synchronisation, suggesting that the process of synchronisation/anti synchronisation cannot be maintained over a long period of time. The variation of the coherence $r_1$ close to the values of +1/-1 for laser frequency of the order $10^{15}$ Hz is shown in fig (3). It shows both synchronisation (+1), anti-synchronisation (-1) as well times for zero coherence. The time for reaching the values +1/-1 is of the order of $10^{-15}$ s. This time interval capability is not currently available in Weihs [5] experiment or any other recent twin photon EPR experiment.

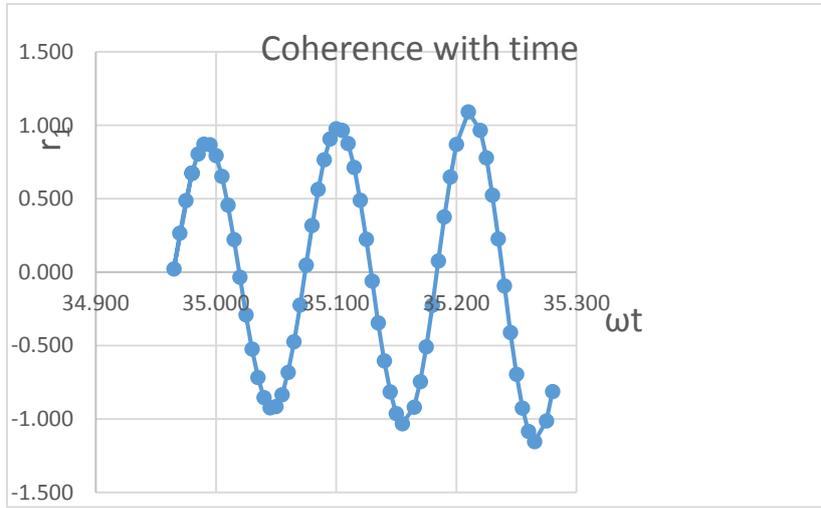

Fig (3)

## 8   Concluding remarks

In this work, a new theory has been used for describing the 'non local' effect observed in EPR. Direction of polarisation of the photons at each arm of the experiment is observed using an independent local measurement. The correlation of the two polarisation directions is attributed to the non local dynamics of the postulated coupled nonlinear oscillators. There are no superluminal signals required to describe this correlation. There is still a question as to whether the non local dynamics of hidden variables in this model, resulting from the localised measurements, can violate Bell's Inequality and therefore be compatible with quantum mechanics.

Using the extended definition of the correlation function of the model, Bell's Inequality can be derived in CHSH [4] form as,

$$|P(\hat{a},\hat{b}) - P(\hat{a},\hat{d})| + |P(\hat{c},\hat{d}) + P(\hat{c},\hat{b})| \leq 2$$

where $\hat{a}, \hat{b}, \hat{c},$ and $\hat{d}$ are the polarisation measurement angles for the usual observers Alice and Bob. The probability of a photon being present at the measurement angle of the analyser can be justifiably taken as the ratio, given by Malus's law for linearly polarised light, of the incident maximum intensity $I_{max}$ to output intensity, $I$ in the direction of measurement, i.e. $\frac{I}{I_{max}} = Cos^2\phi$, where $\phi$ is the angle between $I_{max}$ direction and the measurement direction of the analyser.

Taking value +1 for correlation coincidences between Alice and Bob observations and -1 for anti-correlation coincidences, we can write correlation $P(\hat{a}, \hat{b})$ as $\left(+1(Cos^2\phi) - 1(Sin^2\phi)\right)$ which equals $Cos2\phi$, where $\phi = \hat{b} - \hat{a}$. The correlation calculated for the various polariser angles $\hat{a}, \hat{b}, \hat{c},$ and $\hat{d}$, shows that the Bell's Inequality is violated. In the case, $\hat{a} = 0, \hat{b} = 22.5, \hat{c} = 45,$ and $\hat{d} = 67.5$, we have

$$|P(0,22.5) - P(0,67.5)| + |P(45,67.5) + P(45,22.5)| = 2\sqrt{2}$$

The violation of Bell's inequality confirms that the model we have proposed confirms a non local effect that is compatible with quantum theoretical results. The non locality, we have resolved in the twin photon EPR, has been achieved without the use of superluminal signals. Bohm [2] demonstration that it is possible to produce a theory that is based on local effects to explain the non local effect in EPR has been shown to be valid in the twin photon EPR using the proposed model.

*now retired